\documentclass[amsmath,amssymb,prc,final,showpacs,twocolumn]{revtex4-2}
 
\usepackage{bm,hyphenat,xspace,xcolor}
\usepackage{graphicx,epsfig}

\newcommand {\mbf}[1]{{\mathbf{#1}}}

\begin{document}

\title {$\Lambda$-deuteron scattering below $\Sigma$-threshold}
  
\author{A.~Deltuva} 
\affiliation
{Institute of Theoretical Physics and Astronomy, 
Vilnius University, Saul\.etekio al. 3, LT-10257 Vilnius, Lithuania
}

\received{2 June  2026} 

\begin{abstract}
  $\Lambda d$ elastic scattering and breakup below the  $\Sigma$ threshold
  is described fully including the virtual coupling to $\Sigma$ channels. Well-converged results are obtained
solving the multichannel scattering equations for  three-body transition operators in momentum-space.
Scattering length, exclusive, semi-inclusive and total elastic and breakup cross sections,  and spin observables
are calculated, aiming to study their dependence on kinematic variables and dynamic ingredients.
The uncertainty due to the hyperon-nucleon interaction model is most sizable. 
The observables and regimes that could help to constrain the hyperon-nucleon force models are identified.
The single-channel approximation of the problem is validated at low and intermediate energies.
\end{abstract}

 \maketitle

\section{Introduction \label{sec:intro}}

The vast majority of hypernuclear few-body system studies concentrated on bound and resonant states
\cite{gibson_comparative_1981,nogga_hypernuclei_2002,belyaev2008three,hiyama2014three,%
gal2014there,afnan2015resonances,Gazda:2015qyt,schafer2020continuum,Le:2022ikc,gibson2022possible,sym14112381,lazauskas:26a};
see Ref.~\cite{HAIDENBAUER26a} for a recent review.
  However, the scattering processes may provide a further insight into the features of hypernuclear interactions,
especially regarding higher angular-momentum states and spin-dependent forces, in a similar way that the 
nucleon-deuteron ($Nd$) scattering significantly supplements the knowledge of the nucleon-nucleon ($NN$)
force obtained from the bound-state properties of light nuclei.
Since the very few available experimental data in the hyperon-nucleon ($YN$) sector are used to 
determine the parameters of the $YN$ potentials, the simplest but nontrivial case for testing those models is offered
by the hyperon-deuteron ($Yd$) system. In particular, it involves the hyperon-neutron force, constrained so far by theoretical
considerations based on bound states only.
Though no high-precision hyperon-deuteron scattering data are available at present, the theoretical
calculation of this process is important also for other reactions, such as kaonic disintegration of the trinucleon or
$Yd$ correlation functions in high-energy heavy-ion collisions. 
For the latter quantity in the $\Lambda$-deuteron ($\Lambda d$) system the most advanced study has recently been 
performed by Kohno and Kamada  using Faddeev-type calculations in the momentum space \cite{kohno:24a,kohno:25a}.
Their calculations, however, included the $\Lambda$-$\Sigma$ coupling  at the two-particle level only. That is,
the Lippmann-Schwinger equation with full $\Lambda$-$\Sigma$ coupling was solved for two-particle transition operators,
but then only the  diagonal $\Lambda$-component of those two-body operators was included into three-particle Faddeev
equations. This approximation leads to a significant simplification of three-particle equations, reducing the four-sector
Hilbert space $\Lambda pn \oplus \Sigma^+nn \oplus \Sigma^0pn \oplus \Sigma^-pp$ to a single sector $\Lambda pn$.
An alternative approach is neglecting the $\Lambda$-$\Sigma$ coupling from the beginning but readjusting the $\Lambda N$
potential parameters to the predictions of full potentials at low energies; it has been used recently in the
calculation of $\Lambda pp$ correlation functions \cite{garrido:24a}.

The aim of the present work is to perform rigorous $\Lambda$-deuteron scattering calculations with the full
$\Lambda$-$\Sigma$ coupling in the energy regime below the $\Sigma$-production threshold. This type of calculations
has been reported in Ref.~\cite{lazauskas:26a} but was limited to the total angular momentum and parity
$J^\Pi = \frac12^+$ searching for resonant states. This is obviously insufficient for the converged scattering
results, such as integrated or differential cross sections in the elastic scattering and breakup.
As the results of Ref.~\cite{lazauskas:26a} showed a strong sensitivity to the $YN$ potential parametrization,
this sensitivity will be studied in detail, along with other dynamic ingredients, such as the $NN$ force and the 
Coulomb force in the  $\Sigma^-pp$ channel. Finally, simplified calculations including the $\Lambda$-$\Sigma$ coupling
at the two-particle level but neglecting in the three-particle equations will be performed, enabling to evaluate
the accuracy of the approach used in Refs.~\cite{kohno:24a,kohno:25a}.

The $\Lambda$-$\Sigma$ coupled-channel calculations are conceptually similar to those performed for the $Nd$ scattering 
with the $\Delta$-isobar excitation \cite{deltuva:15d} and for the three-cluster nuclear reactions with core
excitation, such as the inelastic or transfer processes in deuteron-nucleus collisions \cite{deltuva:23a,deltuva:23b}.
The difference is the larger number of coupled Hilbert sectors, four in $\Lambda d$
versus only two in  $Nd$ or nuclear reactions. Nevertheless, the previous developments in $Nd$ scattering
and nuclear reactions with core excitation serve as a good starting point for the rigorous three-body coupled-channel
description of the $\Lambda d$ scattering.

Section II introduces the multichannel three-body scattering equations in an extended Hilbert space.
Section III presents results, while
discussion and conclusions are collected in Sec. IV.

\section{Three-body multichannel scattering equations \label{sec:eq}}

I consider the $YNN$ three-particle system in its center-of-mass (c.m.) frame.
Since the isospin symmetry is only approximate, it is not used, protons and neutron are treated as distinguishable
particles, taking into account also their mass difference.
The hyperon $Y$ can be  in any of the states $\Lambda, \Sigma^+, \Sigma^0, \Sigma^-$,
with respective components of the operators being indicated 
by Latin subscripts in the following. Thus, I have to work in a Hilbert space with four sectors corresponding
to four different internal hyperon states.
The transitions between the internal hyperon states are
induced by hyperon-nucleon  potentials $V^{\alpha}$ with components $V^{\alpha}_{ba}$,
where the Greek superscript $\alpha$
labels the spectator particle in the odd-man-out notation, e.g., spectator $1$ means
the interacting  pair (23), and so on. The free and full Hamiltonians in the Hilbert space with the four sectors are
\begin{subequations}  \label{eq:H}
\begin{align}
      H_0 = & {} K + \Delta M , \\
 H = & {} H_0 + \sum_{\alpha=1}^3 V^\alpha,
\end{align}
\end{subequations}
respectively, where $K$ is the kinetic energy operator and $\Delta M$ the internal Hamiltonian for three particles,
both being diagonal in the Hilbert sectors with respective projections
$K_a$ and  $\Delta m_a$, 
{resulting in $(H_0)_a = K_a + \Delta m_a$.
The energy for internal particle states is measured with respect to the lowest threshold where
the hyperon is $\Lambda$. The eigenvalues of $\Delta M$ are
$\Delta m_a =0$, $[M(\Sigma^+nn)-M(\Lambda np)]$, $[M(\Sigma^0np)-M(\Lambda np)]$, $[M(\Sigma^-pp)-M(\Lambda np)]$
for the $\Lambda, \Sigma^+, \Sigma^0, \Sigma^-$ channels, respectively, where $M(YNN)$ denotes the total mass of the three involved
particles $YNN$, e.g., $M(\Lambda np) = m_\Lambda + m_n + m_p$, $M(\Sigma^+nn) = m_{\Sigma^+} + 2m_n$, and so on.
}
When the hyperon is spectator,
 $V^Y = V_{np}$ is simply the neutron-proton ($np$) potential.
In contrast, the potentials $V^\alpha$ with the neutron or proton being the spectator
dynamically couple three of the four sectors, i.e.,
\begin{subequations}  \label{eq:V}
  \begin{align} \label{eq:Va}
  V^n = {} &
  \begin{pmatrix} V_{\Lambda p,\Lambda p} & V_{\Lambda p,\Sigma^+ n} & V_{\Lambda p,\Sigma^0 p} & 0  \\
V_{\Sigma^+ n,\Lambda p} & V_{\Sigma^+ n,\Sigma^+ n} & V_{\Sigma^+ n,\Sigma^0 p}  & 0 \\
V_{\Sigma^0 p,\Lambda p} & V_{\Sigma^0 p,\Sigma^+ n} & V_{\Sigma^0 p,\Sigma^0 p}   & 0 \\
  0 & 0 & 0 & 0  \end{pmatrix}, \\  \label{eq:Vb}
  V^p = {} &
  \begin{pmatrix} V_{\Lambda n,\Lambda n} & 0 & V_{\Lambda n,\Sigma^0 n} & V_{\Lambda n,\Sigma^- p} \\
  0 & 0 & 0 & 0   \\
V_{\Sigma^0 n,\Lambda n} & 0 & V_{\Sigma^0 n,\Sigma^0 n} & V_{\Sigma^0 n,\Sigma^- p} \\
V_{\Sigma^- p,\Lambda n} & 0 & V_{\Sigma^- p,\Sigma^0 n} & V_{\Sigma^- p,\Sigma^- p}  \end{pmatrix};
  \end{align}
\end{subequations}
for clarity, the subscripts indicate interacting particles in the final and initial states. 

Thus, the projection of the full Hamiltonian to fixed final and initial sectors is
\begin{equation}  \label{eq:Hab}
  H_{ba} = \delta_{ba}(H_0)_a + \sum_{\alpha=1}^3 V^\alpha_{ba}.
\end{equation}

For each interacting pair the two-particle transition operator is obtained
from the Lippmann-Schwinger equation 
\begin{equation}  \label{eq:Tg}
T^{\alpha}_{ba} =  V^{\alpha}_{ba} +\sum_{c} 
V^{\alpha}_{bc} (G_0)_{c} T^{\alpha}_{ca}
\end{equation}
where $c$ runs over $\Lambda, \Sigma^+, \Sigma^0, \Sigma^-$ and 
$(G_0)_{c} =(E+i0-H_0)_{c}^{-1}= (E+i0-\Delta m_c - K_c)^{-1}$ is the free resolvent in the channel $c$ at the available
three-body energy $E$. The two-nucleon transition operator $T^Y_{ba}$ is again diagonal with respect to the four Hilbert sectors,
while the hyperon-nucleon transition operators have the structure
{
\begin{subequations}  \label{eq:T}
  \begin{align}  \label{eq:Ta}
  T^n = {} &
  \begin{pmatrix} T_{\Lambda p,\Lambda p} & T_{\Lambda p,\Sigma^+ n} & T_{\Lambda p,\Sigma^0 p} & 0 \\
T_{\Sigma^+ n,\Lambda p} & T_{\Sigma^+ n,\Sigma^+ n} & T_{\Sigma^+ n,\Sigma^0 p} & 0 \\
T_{\Sigma^0 p,\Lambda p} & T_{\Sigma^0 p,\Sigma^+ n} & T_{\Sigma^0 p,\Sigma^0 p}  & 0 \\
  0 & 0 & 0 & 0  \end{pmatrix}  \\  = {} &    \label{eq:Tb}
  \begin{pmatrix} T^n_{\Lambda\Lambda } & T^n_{\Lambda\Sigma^+ } & T^n_{\Lambda\Sigma^0 } & 0 \\
T^n_{\Sigma^+\Lambda } & T^n_{\Sigma^+\Sigma^+ } & T^n_{\Sigma^+\Sigma^0 } & 0 \\
T^n_{\Sigma^0\Lambda } & T^n_{\Sigma^0\Sigma^+ } & T^n_{\Sigma^0\Sigma^0 }  & 0 \\
  0 & 0 & 0 & 0   \end{pmatrix}, \\  \label{eq:Tc}
  T^p = {} &
  \begin{pmatrix} T^p_{\Lambda\Lambda} & 0 & T^p_{\Lambda\Sigma^0} & T^p_{\Lambda\Sigma^-}  \\
      0 & 0 & 0 & 0  \\
T^p_{\Sigma^0\Lambda}  & 0 & T^p_{\Sigma^0\Sigma^0} & T^p_{\Sigma^0\Sigma^-} \\
T^p_{\Sigma^-\Lambda}  & 0 & T^p_{\Sigma^-\Sigma^0} & T^p_{\Sigma^-\Sigma^-}  \end{pmatrix}.
  \end{align}
\end{subequations}
Sectors $\Sigma^+$ and $\Sigma^-$ do not couple directly via two-particle potential or transition operator.
Equation (\ref{eq:Ta}) is introduced in order to relate the multi-channel transition operator and the respective potential (\ref{eq:Va}),
while the notation (\ref{eq:Tb}) is more convenient in three-particle equations. For brevity only the latter form is given for $T^p$ in
Eq.~(\ref{eq:Tc}).
}

Dynamic equations to be solved  in the present work are the
Faddeev equations  in the Alt-Grassberger-Sandhas (AGS) version  \cite{ags_equations67}
for three-body transition operators $U^{\beta \alpha}_{ba}$, extended  in Ref.~\cite{deltuva:13d} to include
the coupling of different sectors in the Hilbert space. The system of AGS integral equations fully couples different spectators
$Y,p,n$ and sectors $\Lambda, \Sigma^+, \Sigma^0, \Sigma^-$, and 
in the compact form is written as 
\begin{equation}  \label{eq:Uba}
U^{\beta \alpha}_{ba}  = \bar{\delta}_{\beta\alpha} \, \delta_{ba} (G_{0})_a^{-1}  +
\sum_{\sigma=1}^3 \, \sum_{c}   \bar{\delta}_{\beta \sigma} \, T^{\sigma}_{bc}  \,
(G_{0})_{c} U^{\sigma \alpha}_{ca},
\end{equation}
where  $\bar{\delta}_{\beta\alpha} = 1 - \delta_{\beta\alpha}$. Taking Eq.~(\ref{eq:Uba}) with $\beta=0$ one obtains
the transition operator for the breakup of the initial pair, $U^{0 \alpha}_{ba}$, that is not directly involved in the
coupled system of equations, but is calculated as a quadrature from the transition operators (\ref{eq:Uba}) with
$\beta = 1,2,3$.

The physical transition amplitudes are determined by the
on-shell matrix elements of $U^{\beta \alpha}_{ba}$ sandwiched between initial and final channel states.
For  $\Lambda d$ collisions the initial channel state $|d \mbf{q}_i \rangle^Y_\Lambda$ is given by the product
of the deuteron wave function and the plane wave with momentum $\mbf{q}_i$ for the $\Lambda d$ relative motion.
The on-shell condition reads $-\epsilon_d + q_i^2/2M_\Lambda = E$, where $\epsilon_d \approx 2.2245$ MeV is the
deuteron binding energy and $M_\Lambda$ is the reduced $\Lambda d$ mass.
Below the $\Sigma^0 d$ threshold at 74.73 MeV the only inelastic channel is the breakup one
$|  \mbf{p} \mbf{q} \rangle^\beta_\Lambda$, characterized by two Jacobi momenta $\mbf{p}, \mbf{q}$
that can be given in any of the three Jacobi configurations (for any spectator particle) $\beta$.
Their magnitudes are restricted by  energy conservation, $p^2/2\mu_\beta + q^2/2M_\beta = E$, with
$\mu_\beta$ and $M_\beta$ being reduced masses for the pair and spectator, respectively.
Making explicit only the dependence on the final momenta,
the elastic and breakup amplitudes for  $\Lambda d$ scattering are
\begin{subequations}  \label{eq:ampl}
  \begin{align}
\mathcal{A}(\mbf{q}) = {} &
    {}^Y_\Lambda\!\langle  d \mbf{q} | U^{YY}_{\Lambda\Lambda}| d \mbf{q}_i \rangle^Y_\Lambda, \\
\mathcal{B}^\beta(\mbf{p}, \mbf{q}) = {} &
    {}^\beta_\Lambda\!\langle  \mbf{p} \mbf{q} | U^{0Y}_{\Lambda\Lambda}|  d \mbf{q}_i \rangle^Y_\Lambda,
  \end{align}
\end{subequations}
respectively. Once the above amplitudes are calculated, the step to scattering observables
is straightforward. For example, the elastic  differential cross section is
\begin{gather}  \label{eq:dsdo}
\frac{d\sigma}{d\Omega} = (2\pi)^4 \, M_\Lambda^2 \,   |\mathcal{A}(\mbf{q})|^2,
\end{gather}
while the one for breakup is obtained by a partial integration of 
\begin{gather}  \label{eq:d6s}
  \frac{d^6\sigma}{d^3\mbf{p} d^3\mbf{q}} = 
  (2\pi)^4 \, \frac{M_\Lambda}{q_i} \,
\delta \left(E-\frac{p^2}{2\mu_\beta} - \frac{q^2}{2M_\beta} \right) \,
|\mathcal{B}^\beta(\mbf{p},\mbf{q})|^2 
\end{gather}
keeping dependence on desired variables. In the present work I will show the semi-inclusive
angular cross section in the c.m. frame for the detected particle $\beta$, i.e.,
\begin{gather}  \label{eq:dsdob}
  \frac{d\sigma}{d\Omega_\beta}
= (2\pi)^4 \,  \frac{M_\Lambda}{q_i} \, \mu_\beta  \, 
\int d^2\mbf{\hat{p}} \, d q \; {p} q^2
|\mathcal{B}^\beta(\mbf{p},\mbf{q})|^2.
\end{gather}
Alternatively, after integration over all angles  the energy $E_\beta=q^2/2m_\beta$ distribution of the selected particle $\beta$
with mass $m_\beta$ is obtained,
that in the c.m. frame reads
\begin{gather}  \label{eq:dsde}
  \frac{d\sigma}{dE_\beta}
= (2\pi)^4 \, \frac{M_\Lambda}{q_i} \, \mu_\beta  \, m_\beta \,
\int d^2\mbf{\hat{p}} \, d^2\mbf{\hat{q}} \; {p} q
|\mathcal{B}^\beta(\mbf{p},\mbf{q})|^2.
\end{gather}

For brevity, the above notation suppresses the dependence on the spin quantum numbers. Therefore 
for unpolarized observables one has to perform a corresponding spin averaging (summation) over the initial
(final) states in Eqs.~(\ref{eq:dsdo}) - (\ref{eq:dsde}).

In practical calculations the amplitudes (\ref{eq:ampl}) are built from their partial-wave components
since the system of AGS equations (\ref{eq:Uba}) is solved  in the momentum-space 
partial-wave representation. As there are three possible spectator particles $\alpha$
and four Hilbert-space sectors with different
hyperonic content $a$, in total twelve sets of base functions
$|p q \{ l [ (LS) j s] \mathcal{S}\}J M \rangle^\alpha_a $ are employed. They are characterized 
by the magnitudes of Jacobi momenta $p$ and $q$, their associated orbital
angular momenta  $L$ and  $l$, pair and spectator spins $S$ and $s$,  the total pair angular momentum
$j$ and the channel spin $\mathcal{S}$, that all are  coupled, building the total angular momentum $J$ with the projection $M$.
Furthermore, for $a=\Sigma^+$ or $\Sigma^-$ the two nucleons are identical, implying that a
proper antisymmetrization of two-neutron or two-proton states has to be performed.
In the present work, the standard set leading to well converged results is given by all states with
$j \le 3$ and $J\le\frac{39}{2}$.

\section{Results  \label{sec:res}}

The $YN$ interaction is far less constrained by the data than the $NN$ one, leading to a visible model dependence
of some observables other than  ground-state features \cite{lazauskas:26a}. To explore this dependence,
the calculations are performed using the most advanced next-to-leading order potential NLO19 \cite{Haidenbauer:2019boi} 
with $\lambda_{YN}=$ 500, 550, 600, 650, and 700 MeV values of the momentum cutoff parameter. The results will be labeled as
NLO19(500), NLO19(550), NLO19(600), NLO19(650), and NLO19(700), respectively. 
The study of the $J^\Pi = \frac12^+$  resonant states \cite{lazauskas:26a} revealed very weak sensitivity to the $NN$ force model, however,
it is not clear whether it persists also in higher angular momentum states. Therefore two high-precision models of the
$NN$ force are used, the first one based on the chiral effective field theory at the next-to-next-to-next-to-leading order
(N$^3$LO) \cite{entem:03a} and the second one based on meson-exchange theory, the charge-dependent Bonn (CDB) potential 
\cite{machleidt:01a}. The Coulomb force is absent except for the $\Sigma^-pp$ channel, which remains energetically closed
in the present study. Nevertheless, it is interesting to evaluate  also this effect, which can be simply performed adding
the screened Coulomb potential; the screening radius $R=10$ fm appears to be fully sufficient and
no renormalization is needed for $\Lambda n p$ observables. Unless specified explicitly, the calculations are performed using
N$^3$LO $NN$ potential without Coulomb force; the results using CDB potential or the  $\Sigma^-p$ and $pp$ Coulomb force are labeled
by CDB or Coul, respectively. Finally, in order to estimate the effect of  $\Sigma NN$ channels, three-body AGS equations (\ref{eq:Uba})
will be solved  neglecting all the $\Sigma$  components of the $YN$ transition operators (\ref{eq:T}), i.e., keeping only
$T^n_{\Lambda\Lambda }$ and $T^p_{\Lambda\Lambda }$. This kind of results will be labeled as w/o $\Sigma$.

\begin{table}[!]
  \caption{Calculated $\Lambda d$ scattering lengths (in units of fm) for different force models. The last two columns display results
    neglecting $\Sigma NN$ channels in three-body scattering equations.
}
\label{tab:a}
\centering
\begin{ruledtabular}
  \begin{tabular}{{l}*{4}{r}} \
   & \multicolumn{2}{c}{full} & \multicolumn{2}{c}{ w/o $\Sigma$} \\ \hline
  & $a({\frac12^+})$  &  $a({\frac32^+})$   & $a({\frac12^+})$  &  $a({\frac32^+})$ \\ \hline
  NLO19(500)  & 19.48 & -45.29  & 20.51 & -33.82  \\
  NLO19(550)  & 19.26 & -34.54  & 20.77 & -24.04  \\
  NLO19(600)  & 19.08 & -22.70  & 20.68 & -16.82  \\
  NLO19(650)  & 18.58 & -17.24  & 19.59 & -15.13  \\
  NLO19(700)  & 18.11 & -14.56  & 18.49 & -13.38  \\
  NLO19(500),CDB  & 18.75 & -33.87  \\
  NLO19(550),CDB  & 19.30 & -29.58  \\
  NLO19(600),CDB  & 19.50 & -20.55  \\
  NLO19(650),CDB  & 19.14 & -16.08  \\
  NLO19(700),CDB  & 18.76 & -14.19  \\
  NLO19(500),Coul  & 19.37 & -47.17  \\
  NLO19(500)[$j\le 2$]  & 19.58 & -44.69  \\
  NLO19(500)[$j\le 4$]  & 19.47 & -45.26  \\
 \end{tabular}
\end{ruledtabular}
\end{table}

The results for  $\Lambda d$ scattering lengths in the doublet ($J^\Pi = \frac12^+$) and quartet ($\frac32^+$)
states are collected in Table \ref{tab:a}.
Both scattering lengths have large absolute values in comparison to the few fm range of the interaction,
signaling the appearance of near-threshold states in the $\Lambda n p$ system:
the weakly bound ${}^3_\Lambda$H for $J^\Pi = \frac12^+$ and the virtual one for $J^\Pi = \frac32^+$, with positive and negative
scattering length, respectively. This is consistent with previous calculations \cite{kohno:24a}.
Remarkably, while the doublet scattering length shows only weak sensitivity to the force model,  the quartet one exhibits sizable model dependence,
especially with respect to the $YN$ potential, with   NLO19(500) and NLO19(700) predictions differing by a factor of 3. However, one has to keep
in mind that such a large sensitivity is due to proximity of the virtual state, and corresponds to a  0.1 MeV shift in its position. 
The sensitivity to the $NN$ potential can be read off from the predictions with the CDB potential instead of N$^3$LO, and
is also visible, though less pronounced. The effect of the Coulomb force in $\Sigma^-pp$ channel is very small,
but still visible in the very sensitive quartet scattering length. On the other hand,  the last two lines
in Table \ref{tab:a}, together with the first one, demonstrate the  perfect convergence achieved with respect to the
included partial waves.

The last two columns in Table \ref{tab:a} show results neglecting all the $\Sigma$  components in three-body equations.
They provide systematically larger (smaller) absolute values of the doublet (quartet) scattering lengths as compared to full
calculations, indicating that in both channels the virtual  $\Sigma$ excitation yields some additional attraction.
However, the effect is not sizable, such that the full results and those w/o $\Sigma$ partially overlap within their
$YN$ force model uncertainties.

Within their model uncertainties the calculated $\Lambda d$ scattering lengths agree with the experimental evaluation 
\cite{hu:epjconf24} from $\Lambda d$ correlation measurement, $20^{+3}_{-3}$ and $-16^{+2}_{-1}$ fm for the doublet and quartet,
respectively.

\begin{figure}[!]
\begin{center}
\includegraphics[scale=0.64]{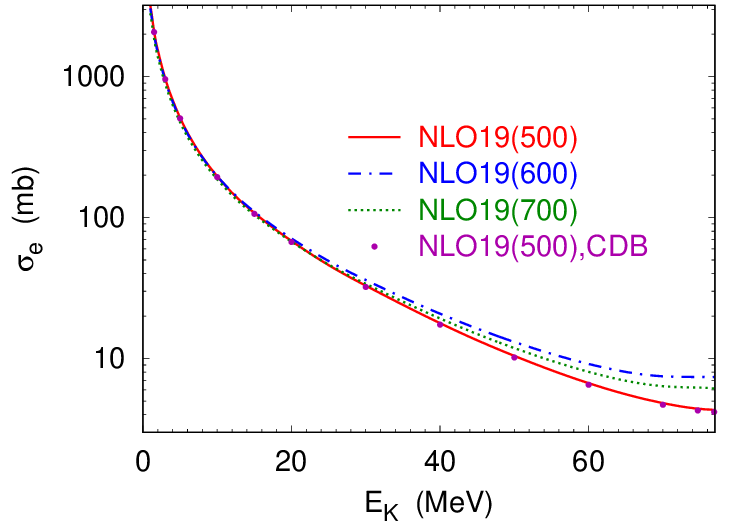}
\end{center}
\caption{\label{fig:el-tot} (Color online)
  $\Lambda d$ total elastic cross section as a function of the kinetic energy in the c.m. frame.
Predictions using different force models are compared.
}
\end{figure}

\begin{figure}[!]
\begin{center}
\includegraphics[scale=0.64]{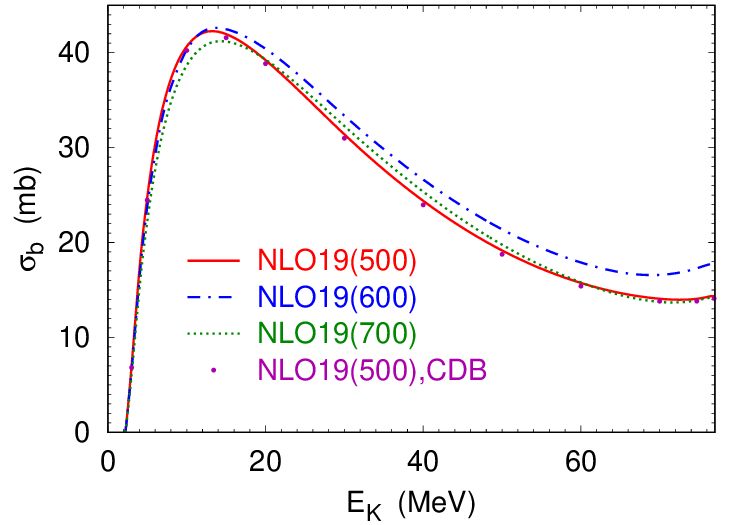}
\end{center}
\caption{\label{fig:br-tot} (Color online)
  $\Lambda d$ total breakup cross section as a function of the kinetic energy in the c.m. frame.
Predictions using different force models are compared.
}
\end{figure}

The total elastic and breakup cross sections for  $\Lambda d$ scattering up to the  $\Sigma^0 d$ threshold is studied in
Figs.~\ref{fig:el-tot} and \ref{fig:br-tot}, respectively.
As there are no experimental data so far, the main focus is on the energy dependence and sensitivity
to $YN$ and $NN$ potentials. As expected from large scattering lengths, the elastic cross section is very large
at the $\Lambda d$ threshold but rapidly decreases with increasing c.m. kinetic energy $E_K = q_i^2/2M_\Lambda$. Away from the
$\Lambda d$ threshold the total elastic cross sections show little sensitivity to the $NN$ potential but
from moderate to large  sensitivity to the $YN$ potential, especially when approaching the $\Sigma^0 d$ threshold.
The total breakup cross section has a typical energy dependence, with a maximum near $E_K=15$ MeV. Again, it shows
little sensitivity to the $NN$ potential but moderately depends  on the used $YN$ potential. Notably, it starts to increase
again when approaching the $\Sigma$ threshold, a behavior seen in the  $J^\Pi = \frac12^+$ partial wave \cite{lazauskas:26a}.
It turns out that this particular state is largely responsible for the observed increase, a probable explanation being the
resonant $\Sigma NN$ state, located in the complex energy plane not far from $\Sigma^0 d$ threshold. As found in
Ref.~\cite{lazauskas:26a}, that resonant state has predominantly isospin 1, and therefore affects mostly the $\Lambda d$
breakup channel but not  the elastic one that has isospin 0.

\begin{figure}[!]
\begin{center}
\includegraphics[scale=0.64]{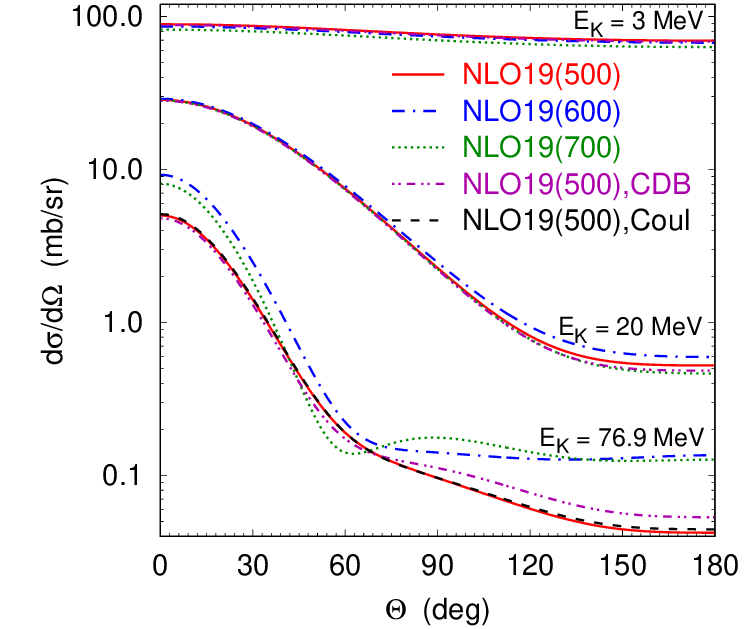}
\end{center}
\caption{\label{fig:el-dcs} (Color online)
   Differential cross section for $\Lambda d$ elastic scattering at 3, 20 and 76.9 MeV c.m. kinetic energy
  as a function of the c.m. scattering angle.
Predictions using different force models are compared.
}
\end{figure}

The elastic differential cross section in the c.m. frame as a function of scattering angle is studied
in Fig.~\ref{fig:el-dcs} at $E_K=3$, 20, and 76.9 MeV energies.
At the lowest energy the $S$-wave scattering dominates, thus, the seen sensitivity to the
$YN$ potential can be traced back to differences in the predicted  quartet scattering  length.
At the intermediate energy the sensitivity to the $YN$ potential shows up moderately at backward angles,
while at the highest energy it is sizable everywhere. This is probably due to the fact that
$\Sigma$-part of the $YN$ is less constrained by the data, and  the proximity of the $\Sigma^0 d$ threshold located at
76.96 MeV enhances the importance of $\Sigma N$ and $\Lambda N - \Sigma N$ interactions. 
{The sensitivity to the $NN$ force model is visible, but well below the one for the $YN$ potential.
The effect of the Coulomb force, present in the closed $\Sigma^-pp$ channel only, was investigated as well and found to be very small.
Except for the vicinity of the $\Sigma$ threshold,  the results including Coulomb are not distinguishable in the plots and therefore are not shown separately.
A tiny Coulomb force effect is shown at 76.9 MeV, where the proximity to the $\Sigma$ threshold enhances the importance of $\Sigma NN$ channels,
together with the Coulomb force in the $\Sigma^-pp$ channel.}

\begin{figure}[!]
\begin{center}
\includegraphics[scale=0.7]{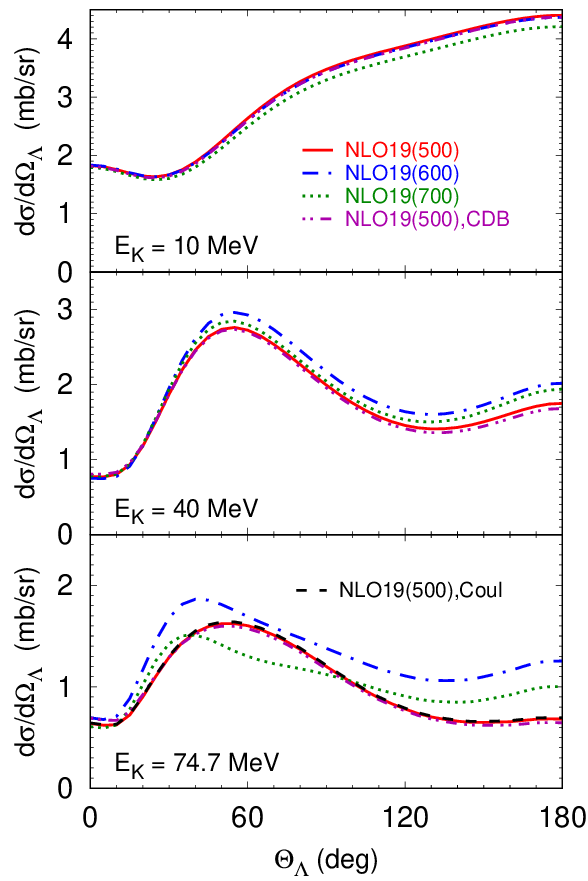}
\end{center}
\caption{\label{fig:br-L} (Color online)
   Semi-inclusive differential cross section for $\Lambda d$ breakup at 10, 40 and 74.7 MeV c.m. kinetic energy
  as a function of the $\Lambda$ scattering angle in c.m. frame.
Predictions using different force models are compared.
}
\end{figure}

\begin{figure}[!]
\begin{center}
\includegraphics[scale=0.7]{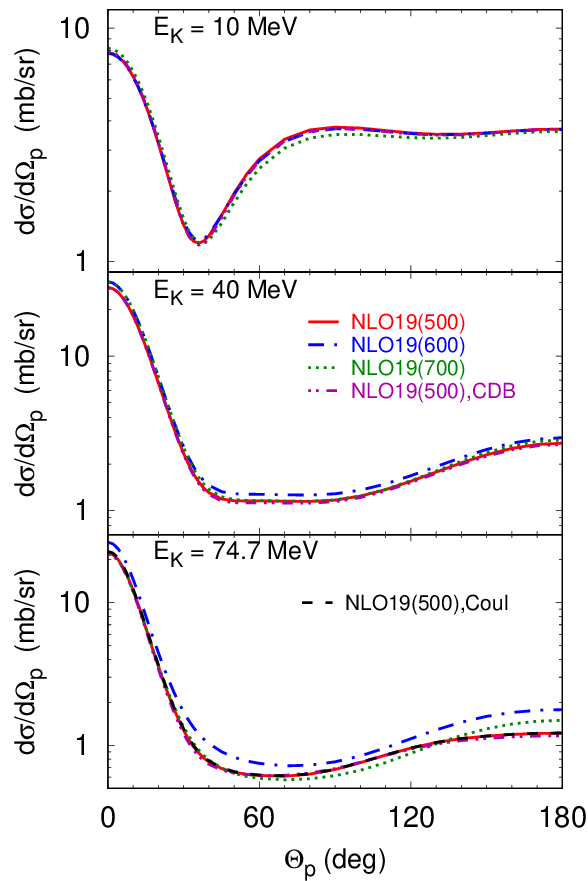}
\end{center}
\caption{\label{fig:br-p} (Color online)
   Semi-inclusive differential cross section for $\Lambda d$ breakup at 10, 40 and 74.7 MeV c.m. kinetic energy
  as a function of the proton scattering angle in c.m. frame.
Predictions using different force models are compared.
}

\end{figure}

\begin{figure}[!]
\begin{center}
\includegraphics[scale=0.64]{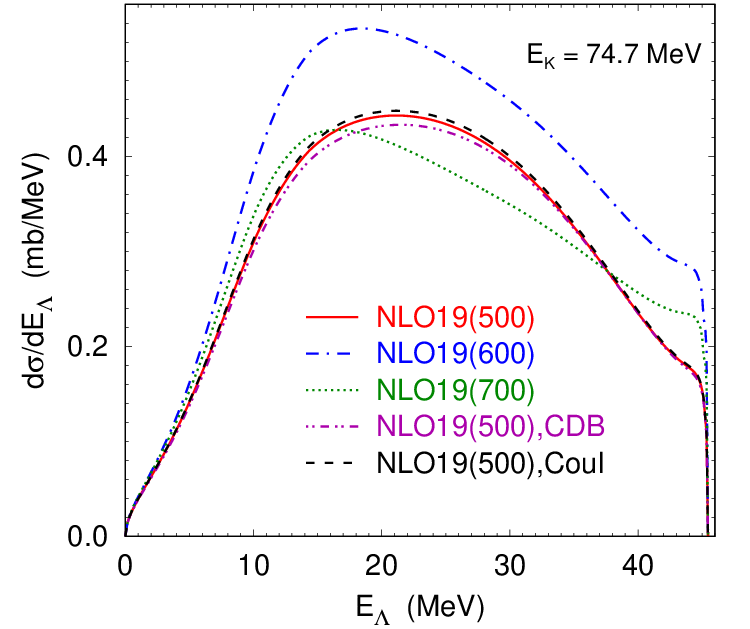}
\end{center}
\caption{\label{fig:br-EL} (Color online)
   Semi-inclusive differential cross section for $\Lambda d$ breakup at  74.7 MeV c.m. kinetic energy
  as a function of the $\Lambda$ energy in c.m. frame.
Predictions using different force models are compared.
}
\end{figure}

\begin{figure}[!]
\begin{center}
\includegraphics[scale=0.64]{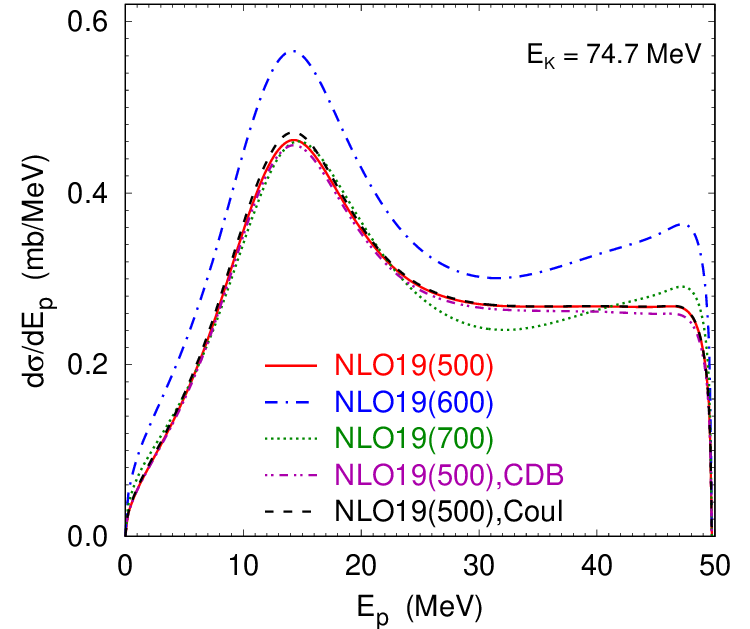}
\end{center}
\caption{\label{fig:br-Ep} (Color online)
   Semi-inclusive differential cross section for $\Lambda d$ breakup at 74.7 MeV c.m. kinetic energy
  as a function of the proton energy in c.m. frame.
Predictions using different force models are compared.
}
\end{figure}

The breakup state is characterized by five continuous variables. Aiming to present a more concise study of the
$\Lambda d$ breakup I therefore show semi-inclusive angular differential cross sections for $\Lambda$ and proton
in Figs.~\ref{fig:br-L} - \ref{fig:br-p}.
Notably, they have very different shape: the proton is preferably emitted at small angles, while $\Lambda$ around
50 deg and at backward angles.
The sensitivity to the chosen $YN$ potential is quite small or moderate at $E_K =10$ and 40 MeV, and only becomes
significant when approaching the $\Sigma$ threshold.
This is confirmed also in the energy distributions shown in Figs.~\ref{fig:br-EL} - \ref{fig:br-Ep}.
Again the shapes for detected hyperon and nucleon 
are different corresponding to different reactions mechanisms. For example, the angular forward peak
and the 14 MeV peak in the proton energy distribution can be explained by the quasi-free $\Lambda n$ scattering
and thus would be particularly well suited for the study of the  $\Lambda n$ interaction that cannot be probed directly.
Figures \ref{fig:br-L} - \ref{fig:br-Ep} show also the results with CDB potential and Coulomb force; as in the elastic
scattering, though discernible, these effects are far less significant than the  $YN$ potential dependence. 
{The Coulomb effect is only discernible near the $\Sigma$ threshold, and therefore is shown at $E_K = 74.7$ MeV only.}


\begin{figure}[!]
\begin{center}
\includegraphics[scale=0.64]{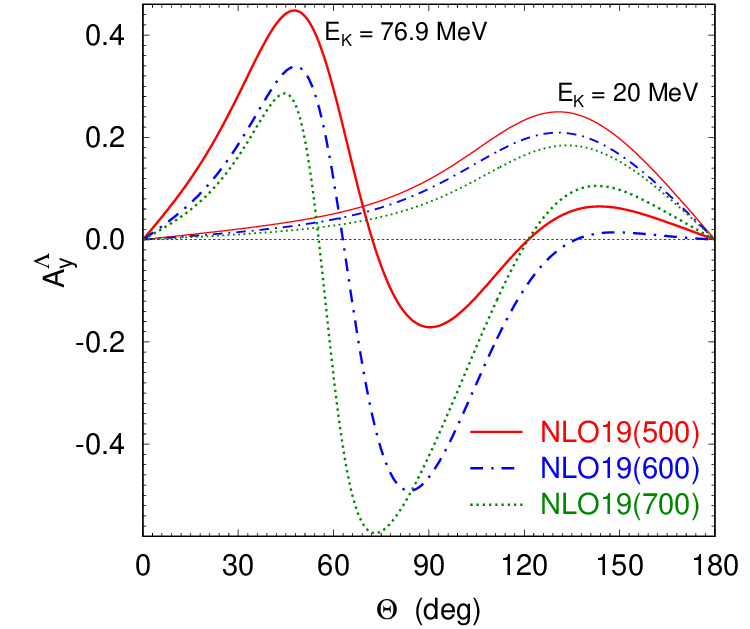}
\end{center}
\caption{\label{fig:el-A} (Color online)
Hyperon analyzing power for $\Lambda d$ elastic scattering at  20 (thin curves) and 76.9 MeV (thick curves) c.m. kinetic energy
  as a function of the c.m. scattering angle.
Predictions using different force models are compared.
}
\end{figure}

Another interesting question is the spin dependence of the interaction, leading to polarization phenomena
characterized by analyzing powers, spin correlation and transfer coefficients. As example I consider
in Fig.~\ref{fig:el-A} the elastic analyzing power of the hyperon, that is equal (up to a sign) to the  polarization
of the outgoing hyperon in the  elastic $\Lambda d$ scattering with unpolarized beam and target. Typically,
analyzing powers are very small at low energies but increase with increasing energy and also change the shape.
The analyzing power of the deuteron, not displayed separately, shows a qualitatively similar energy and angular dependence.
Those features resemble quite closely also polarization observable behavior in the nucleon-deuteron scattering.

\begin{figure}[!]
\begin{center}
\includegraphics[scale=0.64]{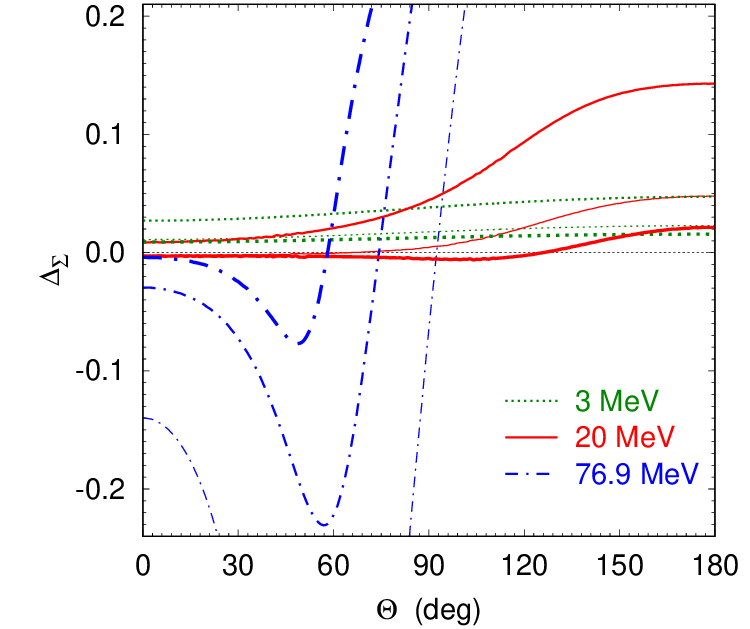}
\end{center}
\caption{\label{fig:el-sdiff} (Color online)
The relative effect of the $\Sigma NN$ channels for the $\Lambda d$ elastic
  differential cross section  at 3, 20 and 76.9 MeV c.m. kinetic energy.
{  The thin, middle, and thick curves correspond to NLO19(500), NLO19(600), and NLO19(700) force models,
  respectively, with the full calculations shown in Fig.~\ref{fig:el-dcs}.}
}
\end{figure}

\begin{figure}[!]
\begin{center}
\includegraphics[scale=0.64]{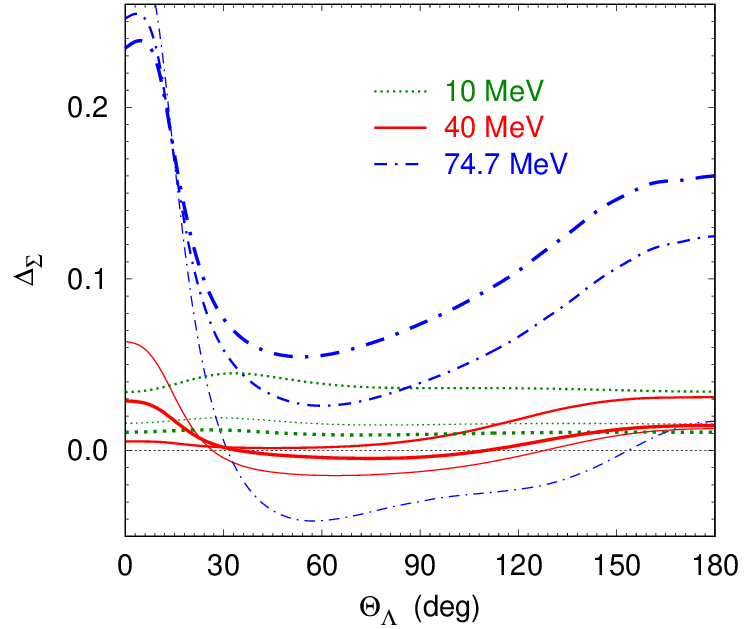}
\end{center}
\caption{\label{fig:b-sdiff} (Color online)
The relative effect of the $\Sigma NN$ channels for the  angular semi-inclusive differential cross section of $\Lambda d$ breakup
at 10, 40 and 74.7 MeV c.m. kinetic energy.
{The thin, middle, and thick curves correspond to NLO19(500), NLO19(600), and NLO19(700) force models,  respectively, 
with the full calculations shown in Fig.~\ref{fig:br-L}.}
}
\end{figure}

Finally I will evaluate the reliability of the approximation  w/o $\Sigma$. I define the relative effect
of  $\Sigma NN$ channels for the observable $\mathcal{O}$ as
\begin{equation}  \label{eq:SNN}
\Delta_\Sigma = \frac{\mathcal{O}(\mathrm{full}) - \mathcal{O}(\mathrm{w/o}\;\Sigma)}{\mathcal{O}(\mathrm{full})}.
\end{equation}
Figures \ref{fig:el-sdiff} - \ref{fig:b-sdiff} display $\Delta_\Sigma$ for differential cross section in elastic
$\Lambda d$ scattering and breakup at several energies using NLO19(500), NLO19(600), and NLO19(700)
force models. The main message is that the effect becomes sizable in the vicinity of the $\Sigma$ threshold,
but is small to moderate at low and intermediate energies. Quantitatively the effect depends also on the
underlying $YN$ potential. 
{ For the breakup differential cross section in the representation of Figs.~\ref{fig:br-p} - \ref{fig:br-Ep} the relative effect
at the  respective energies is of comparable magnitude as presented in Fig.~\ref{fig:b-sdiff} and therefore is not shown separately.
Notably, near the $\Sigma$-threshold   $\Delta_\Sigma$ in the elastic scattering appears to be more sizable than in
the breakup. The scale of Fig.~\ref{fig:el-sdiff} is chosen to represent well the cases of $E_K = 3$ and 20 MeV where the 
approximation  w/o $\Sigma$ can be expected to be reasonable. On the contrary, 
elastic $\Delta_\Sigma$ at $E_K= 76.9$ MeV for all three models exceeds the scale of Fig.~\ref{fig:el-sdiff} 
and reaches 0.9 at backward angles. A possible explanation is that the angular regime with very small elastic differential cross section 
is most sensitive to the details of the underlying interaction. This appears to be in a close analogy with the nucleon-deuteron scattering where
the virtual $\Delta$-isobar excitation becomes most important around the minimum of the elastic differential cross section \cite{deltuva:15d}.
}

\section{Discussion and conclusions \label{sec:sum}}

$\Lambda d$ elastic scattering and breakup in the energy regime up to the $\Sigma$ threshold was described
fully including the coupling to $\Sigma^+nn$, $\Sigma^0pn$, and $\Sigma^-pp$ channels. Integral equations
for multichannel three-body transition operators were solved in the momentum-space partial-wave representation.
High-precision $NN$ potentials and NLO19 models of $YN$ potentials were used. 

Doublet and quartet scattering lengths, total elastic and breakup cross sections, elastic differential cross section and analyzing
powers, and semi-inclusive differential breakup cross sections were calculated, aiming to study their dependence on
kinematic variables and dynamic ingredients.
The sensitivity to the $YN$ potential is dominating, 
it is most sizable for all observables in the regime near the $\Sigma$ threshold, where the virtual
$\Sigma$ excitations are expected to play the most important role. 
Another case is the quartet scattering length and the low-energy elastic cross section, where the sensitivity to
all dynamic ingredients is strongly enhanced due to the presence of near-threshold virtual state pole.

The notable features of the calculated observables are
(i) the increase of the breakup cross section in the vicinity of the $\Sigma$ threshold,
(ii) the characteristic distributions of hyperons and nucleons in the breakup channel, in particular, the peak structures
related to the quasi-free $\Lambda n$ scattering,
particularly well suited for the study of the  $\Lambda n$ interaction, and
(iii) large values of analyzing powers or polarizations enabling the study of the spin-dependent forces.

The single-channel approximation of the problem, i.e., keeping  $\Sigma$ channels in the potential
but neglecting in three-body equations, was found to be a reasonable approximation except near the $\Sigma$ threshold.
Otherwise, the uncertainty caused by this approximation does not exceed the model dependence of the $YN$ potential,
such that the two groups of results overlap at least partially. Since this observation is based on a broad spectrum of
differential cross sections derived from half-shell matrix elements of AGS transition operators, it 
validates with confidence also the calculations of $\Lambda d$ correlation functions
 by Kohno and Kamada \cite{kohno:24a,kohno:25a} relying on the above approximation. 

 The developments of the present work will also serve as a starting point for the description of hypernuclear three-body systems
 above the $\Sigma$ threshold with a large variety of open channels.

 \vspace{1mm}
 The author thanks R. Lazauskas, D. Gazda, and M. Sch\"afer for discussions.
This work has received funding from the 
Research Council of Lithuania (LMTLT) under Contract No.~S-MIP-25-15.
Part of the computations was performed using the infrastructure of the Lithuanian Particle Physics Consortium 
in the framework of agreement No. VS-13 of Vilnius University with LMTLT.


\end{document}